\documentstyle{mn}
\input{epsf}

\title[Redshift distribution of Ly-${\bf \alpha}$ lines and metal systems]
{Redshift distribution of {\bf Ly-$\alpha$} lines and metal systems}

\author[Demia\'nski et al.]
       {M. Demia\'nski$^{1,2}$,  A.G. Doroshkevich$^{3,4}$,
	\& V.Turchaninov$^4$\\
	$1$Institute of Theoretical Physics,
                       University of Warsaw,
                       00-681 Warsaw, Poland\\
        $2$Department of Astronomy, Williams College,
           Williamstown, MA 01267, USA\\
	$3$Theoretical Astrophysics Center,
          Juliane Maries Vej 30,
          DK-2100 Copenhagen \O, Denmark\\
	$4$Keldysh Institute of Applied Mathematics,
                        Russian Academy of Sciences,
                        125047 Moscow,  Russia\\
}

\date{Accepted ...,
     Received 1999 ... ;
in original form 1999 ...  }
\begin{document}
\maketitle

\begin{abstract}
The observed redshift distribution of Ly-$\alpha$ lines and metal
systems is examined in order to discriminate and to trace the evolution
of structure elements observed in the galaxy distribution, at small
redshifts, and to test the theoretical description of structure
evolution. We show that the expected evolution of filamentary component
of structure describes quite well the redshift distribution of metal
systems and stronger Ly-$\alpha$ lines with $\log(N_{HI})\geq$14, at
$z\leq$ 3. The redshift distribution of weaker Ly-$\alpha$ lines
can be attributed to the population of poorer structure elements
(Zel'dovich pancakes), which were formed at high redshifts from 
the invisible DM and non luminous baryonic matter, and at lower 
redshifts they mainly merged and dispersed.

\end{abstract}


\section{Introduction}

During the last years an essential progress has been achieved
both in the observations and interpretations of absorption spectra of
quasars. Now a representative set of absorption lines with
measured redshifts, column densities, and Doppler parameters is
available and the main characteristics of absorbers and their
evolution are, in general, established (see, e.g., Carswell 1995;
Cristiani et al. 1995, 1996; Hu et al. 1995; Fernandez-Soto et al.
1996; Ulmer 1996; Cooke et al 1997; Dinshaw et al. 1995; Bahcall
et al. 1993, 1996; Kim et al. 1997). The available information
suggests various descriptions of absorbers ranging from
pressure-confined gaseous clouds to associations with halos of
galaxies (see discussions in Rees 1995; Charltone 1995;
Fernandes-Soto et al. 1996; Miralda-Escude et al. 1996;
Petitjean 1997). If at low redshifts a significant number of
stronger Ly-$\alpha$ lines and metal systems are certainly
associated with galaxies (Bergeron et al. 1992; Lanzetta et al.
1995; Cowie et al. 1995; Tytler et al. 1995; Le Brune et al. 1996)
than the population of weaker absorbers, dominating at higher redshifts,
has practically disappeared by the redshift $z\approx$ 2  and only
some  weak Ly-$\alpha$ lines are observed far from galaxies even 
at small redshifts (Morris et al. 1993; Shull 1997).

Now the origin and evolution of absorbers are linked to the 
process of formation and evolution of high density structure 
elements (Petitjean et al. 1995; Miralda-Escude et al. 1996; 
Hernquist et al. 1996; Zhang et al. 1997, 1998; Theuns et al. 
1998, 1999; Bryan et al. 1999; Dav\`e et al. 1999; Weinberg et al. 
1998; Machacek et al. 2000). The numerical simulations 
of  dynamical and thermal evolution of gaseous component use 
the same  spectrum of initial perturbations that is responsible 
for the formation   of DM and observed galaxy spatial distributions, 
and they successfully reproduce  the main observed properties
of absorbers. This allows us to consider the evolution of absorbers 
in a context of a more general process of nonlinear evolution of 
small initial  perturbations of DM component and formation of 
observed galaxy  distribution and to compare theoretical
conclusions  with observations. 

This approach links the properties of absorbers, at high redshifts, 
with the spatial galaxy distribution at small redshifts and allows 
to trace some observational characteristics of nonlinear evolution 
of matter distribution up to redshifts $z\sim$3. 

The analysis of large modern redshift surveys such as
the Durham/UKST Galaxy Redshift Survey (Ratcliffe et al., 1996)
and  the Las Campanas Redshift Survey (Shectman et al., 1996)
demonstrates that the galaxy distribution in the universe
can be roughly described as a joint network composed of galaxy 
filaments and walls. The analysis of both surveys shows that
about 50\% of galaxies are concentrated in wall-like
elements surrounding huge underdense regions with
a typical size of 50 -- 100$h^{-1}$ Mpc ($h=H_0/100$ km/s/Mpc
is the dimensionless Hubble parameter) (Doroshkevich et al. 1996,
hereafter LCRS1; Doroshkevich et al. 2000).
In many respects such elements are similar to the Great Wall
(de Lapparent, Geller \& Huchra 1988; Ramella, Geller, \& Huchra
1992). Galaxy filaments appear between walls and
intertwine them into a joint structure. The hierarchy of poorer
and sparser filaments -- including  single galaxies -- can be
considered as a bridge between the richer and poorer structure
elements formed by the non luminous baryonic and dark matter (DM). 
Such poorer elements contain some fraction of hot gas and could 
be seen as Ly-$\alpha$ absorbers away from galaxies. Observations
of weak Ly-$\alpha$ absorbers in voids, mentioned above, can be
attributed to those poorer DM structure elements.

The evolution of structure elements formed by dark matter was
investigated both theoretically, using the Zel'dovich theory
(e.g., Zel'dovich 1970; Zel'dovich \& Novikov 1983; Shandarin 
\& Zel'dovich 1989), and numerically (e.g., Shandarin et al. 
1995; Cole et al. 1997, 1998; Jenkins et al. 1998; Doroshkevich 
et al. 1999, hereafter DMRT; Demia\'nski et al. 2000, hereafter 
DDMT). The statistical description of DM structure evolution 
for the CDM-like power spectra was given in Demia\'nski ~\&~ 
Doroshkevich (1999 hereafter DD99) and in DDMT. The 
main simulated characteristics of walls are found to be consistent 
with theoretical expectations.  

Some progress in the theoretical interpretation of properties of absorbers
has been recently achieved by Hui et al. (1997) and Nath (1997),
who used the Zel'dovich approximation for a semi analytical 
description of some of the absorbers characteristics. 
Another considered theoretical approaches were based on the 'minihalo' 
model (Meiksin 1994), 'Cosmic Webb' theory (Bond \& Wadsley 1997) and 
Press-Schechter method (Valageas et al. 1999). 

In this paper we examine the observed redshift distribution
of absorbers using the available information about the structure
parameters at small redshifts (LCRS1; Doroshkevich et al. 2000; 
DMRT) and statistical description of DM structure evolution 
(DD99, DDMT) in models with CDM-like power spectra. We assume 
that the Lyman-$\alpha$ clouds trace the potential wells formed 
by DM pancakes, filaments and walls. This assumption was widely 
discussed earlier (Rees 1986; Ikeuchi \& Ostriker 1986; Bond, 
Szalay \& Silk 1988; Miralda-Escude ~\&~ Rees 1993; Meiksin 1994; 
Hui et al. 1997; Nath 1997; Bond \& Wadsley 1997; Valageas et al. 
1999), and it is qualitatively consistent with results of numerical 
simulations (Petitjean et al. 1995; Hernquist et al. 1996; 
Petitjean 1997; Theuns et al. 1997; Dav\`e et al. 1999; Weinberg 
et al. 1998). 

The potential of this approach is limited due to the small number of 
observed absorbers and because the characteristics of structure 
derived from observations are not sufficiently accurately known. 
The theoretical description of DM structure evolution is also not 
complete. In particular, it cannot yet describe adequately the 
important process of disruption of structure elements and their 
transformation into a system of high density clumps. This process 
depends on  many factors (Doroshkevich 1980; Vishniac 1983; Valinia 
et al. 1997) and strongly influences the observed properties
of  absorbers both at large and small redshifts (Miralda-Escude 
et al. 1996; DMRT; DDMT). Moreover, the evolution of DM structure 
elements and observed absorbers is not identical. 

Nonetheless, even so limited information allows us to obtain a 
reasonable description and interpretation of some of the observed 
characteristics of absorbers. In this paper we consider the 
statistical description of redshift distribution of absorbers 
and metal systems which is more detailed, in some respects, then 
that usually used. The comparison of measured redshift distributions 
with approximate theoretical expectations (DD99) allows us to 
discriminate statistically the contributions of high density 
filamentary and lower density sheet-like structure elements 
formed by DM and gaseous components of matter, and to trace their 
evolution. These two components of structure elements can be 
roughly identified with two subpopulations of absorbers discussed, 
for example, in Bahcall et al. (1996) and Weymann et al. (1998). 

This division  of structure elements into filaments and more 
sheet-like walls and pancakes is inevitably statistical only. 
The more detailed analysis of DM and galaxy distribution, at small 
redshifts (DMRT, DDMT), demonstrates limitations of such description 
and shows that usually the morphology of structure elements can be 
more accurately characterized in terms of {\it degree of filamentarity 
and/or sheetness} only. The investigation of morphology and spatial 
distribution of DM and gaseous structure elements in simulations 
at high redshifts with the new powerful techniques (such as the 
Minimal Spanning Tree or Minkowski Functional), as well as direct 
comparison of morphological and other characteristics of the structure 
elements with their redshift distribution is required to clarify this 
problem and to obtain more detailed comparison of observed and simulated 
matter distributions. 

This paper is organized as follows. The theoretical model of
structure evolution and the technique necessary for the analysis
are briefly described in Sec. 2. Sec. 3 contains information
about the used observational databases. The results of statistical
analysis are presented in Sec. 4. Discussion and conclusion
can be found in Sec. 5.

\section{ Redshift distribution of absorbers}

\subsection{Pancakes, filaments and walls associated with absorbers}

The redshift distribution of absorbers can be characterized
by the mean comoving free path between absorbers $<l>$
(Sargent et al. 1980) as
$$dN(z) = <l>^{-1}c(1+z)dt = <l>^{-1}{c~|dz|\over H(z)}\eqno(2.1)$$
$$H(z)= H_0\sqrt{\Omega_m(1+z)^3+
(1-\Omega_m-\Omega_\Lambda)(1+z)^2+\Omega_\Lambda}$$
where $dN(z)$ is the number of absorbers between $z$ and $z+dz$,
$<l>^{-1}$ is the mean comoving linear number
density of absorbers, $c$ is the speed of light, $\Omega_m$
and $\Omega_\Lambda$ are correspondingly the dimensionless matter 
density and cosmological constant. The redshift dependence of 
the mean comoving linear number density of absorbers can be 
written as:
$$n_{abs}={c\over H_0}<l>^{-1} = n_w(z) + n_f(z) +
n_{pan}(z),\eqno(2.2)$$
$$dN(z) = n_{abs}{H_0~dz\over H(z)},$$
where the three terms in (2.2) correspond to three main types of
structure elements observed at small redshifts: the rich walls
($n_w$),  filaments ($n_f$), formed by galaxies, and poor
pancakes ($n_{pan}$), situated far from observed galaxies.

Such division of structure elements into three different 
subpopulations is made for the sake of simplicity only. In 
particular, there is an anisotropic sheet-like low density halo 
around all filaments, which in turn are later strongly disrupted 
into a system of high density clumps. The higher density ridges 
cross often the low mass pancakes. Walls are also often formed 
by a system of high density clumps and filaments connected by 
lower density bridges. This means that in nature, there is 
really only one population of structure elements with a broad 
variety of continually distributed  properties, and these elements 
can be considered as the more typical and conspicuous ones. 
Nonetheless, the considered classification  makes this analysis 
simpler and more transparent, as the rate of structure evolution 
depends upon the degree of filamentarity or sheetness.

\subsection{Evolution of DM structure elements}

In the DM dominated universe the basic characteristics of matter 
distribution are determined by the process of formation and evolution 
of DM structure elements. The observed characteristics of absorbers 
and parameters of DM structure elements are not identical but they 
are closely connected and relations between these parameters can 
emerge from suitable investigations.

For the CDM-like transfer function (Bardeen et al. 1986) and
the Harrison--Zel'dovich initial power spectrum the approximate
statistical description of the formation and evolution of DM
structure elements based on the statistics of initial
perturbations was obtained in DD99. The random formation and
merging of Zel'dovich pancakes with 'mass' (column density of a
pancake) $m$ and their evolution, due to the transversal expansion
or compression, are described by approximate expressions similar
to the Press-Schechter relations. Similar approximate relations 
can be also written for the filamentary component of the structure. 
Here we will apply some of these results to describe the observed 
evolution of absorbers.

As is usual in the Zel'dovich theory the redshift dependence of
all characteristics of structure is expressed through a function
$B(\Omega_m,z)$, which describes the growth of perturbations in
the linear theory. For the flat universe with $\Omega_m+\Omega_\Lambda
= 1,~~\Omega_m\geq 0.1$, the function $B(z)$ can be approximated (DD99)
with a precision better then 10\% by the expression
$$B(z)^{-3}\approx {1-\Omega_m+2.2\Omega_m(1+z)^3\over
1+1.2\Omega_m},\eqno(2.3a)$$
and for an open universe with $0.1\leq \Omega_m\leq 1,~ \Omega_
\Lambda=0$ as
$$B(z)^{-1}\approx 1+{2.5\Omega_m\over 1+1.5\Omega_m}z\eqno(2.3b)$$
(Zel'dovich \& Novikov 1983). For $\Omega_m=1,~~\Omega_\Lambda=0$
both expressions give $B^{-1}(z)=1+z$.

\subsubsection{Formation and evolution of DM structure}

The evolution of DM structure can be outlined as a progressive
matter concentration within more and more massive structure elements
which can be roughly discriminated into pancakes, filaments and walls.
The structure elements are primarily formed as Zel'dovich pancakes 
(Shandarin et al. 1995), but later are successively transformed into 
filaments and/or (elliptical) clouds. At high redshifts $z\geq$ 5 -- 
7 a significant fraction of matter ($\sim$80\%) can be compressed 
into low mass pancakes with
$M_{cl}\sim~~ 10^7$ -- $10^9 M_\odot$. Later some of such clouds
can evolve into dwarf galaxies, some could disperse, but the
main fraction will be accumulated by more massive structure elements. 
Properties of the surviving and relaxed clouds can be similar to those
discussed in the 'minihalo' model (Rees 1986, 1995; Miralda-Escude
\& Rees 1993; Meiksin 1994)

The filamentary component of structure also forms at high
redshifts due to the compression of some part of pancakes, and, at
$z\sim$ 1 -- 3, it is, probably, the most conspicuous population of
high overdensity structure elements accumulating 40 -- 50\% of DM
particles (see., e.g., Governato et al. 1998, Jenkins et al. 1998).
Further evolution of structure can be described as a successive
merging of lower mass pancakes and filaments into more and more
massive wall-like structure elements. The largest observed wall-like 
elements, similar to the Great Wall, are formed at redshifts $z\leq$1 
due to the infall and intermixture of smaller pancakes and filaments. 

In observations, at $z\leq$ 1, galaxy filaments are usually situated
within extended under dense regions surrounded by higher
density walls, and they form an irregular broken network. 
This network can be described as a trunk, that is the longest filament
of the complex, and a set of shorter branches. This means that the
spatial distribution of filaments can be conveniently characterized  
by the mean 2D surface number density, $\sigma_f(z)$, that is the mean
number of filaments intersecting a unit area of arbitrary orientation.
For observed galaxy filaments it is estimated as $\sigma_f(0)\approx
(1 - 0.7)\cdot10^{-2}h^2{\rm Mpc}^{-2}$ (LCRS1). The spatial 
distribution of observed walls can be characterized by their mean 
separation $\langle D_w\rangle$ and by their linear density, $\sigma_w$, 
that is the mean number of walls intersecting a unit length of a 
random straight line. For the observed galaxy walls it is estimated 
as $\langle D_w\rangle\approx$40 -- 60$h^{-1}$Mpc that corresponds 
to $\sigma_w\approx$2.5 -- 1.6$\cdot 10^{-2}h$Mpc$^{-1}$. The low 
mass pancakes can be considered as isolated objects and characterized 
by the mean 3D number density $\nu_{pan}(z)$.

Simulations show that rich pancakes and filaments are relaxed (at
least along shorter axes) and are surrounded by an extended DM (and
gaseous) halo. Central parts of dense elements are gravitationally
confined and are unstable with respect to the small scale disruption
(DDMT). Some of the high density clumps are usually embedded in 
filaments and walls what essentially accelerates the relaxation of 
compressed matter. In observations, effects of such disruptions are 
seen as groups and clusters of galaxies. 

The basic properties of DM structure elements vary in a broad range,
but it can be expected that galaxies are formed mainly within
richer high density filaments and pancakes. The essential fraction 
of poor pancakes and, perhaps, filaments, can be invisible, as it 
does not contain sufficiently bright galaxies. Such structure elements 
can be observed as weak Ly-$\alpha$ absorbers situated far from any 
galaxy. Some number of such absorbers situated at a
distance of up to 5$h^{-1}$Mpc from the closest galaxy was found 
by Morris et al. (1993) and Shull (1997). 

This description agrees well with results of numerical simulations 
at small redshifts (DDMT). The validity of this description, when 
it is applied to structure at high redshifts, is not yet reliably 
verified with available simulations due to a small density contrast 
of poor structure elements. The first tests show, however, the 
existence of three kinds of structure elements, namely, high density 
filaments and clumps, and low density pancakes in DM spatial 
distribution at $z=$3. The theoretical description (DD99) confirms 
also the self-similar character of structure evolution (at least 
when the Zel'dovich approximation can be applied). This picture 
is also qualitatively consistent with other simulations of 
structure evolution (Miralda-Escude et al. 1996; Governato et 
al. 1998; Jenkins et al. 1998; Meiksin 1998; DMRT; Dav\`e et 
al. 1999; Weinberg et al. 1998). 

\subsubsection{Main factors of structure evolution}

The four main factors characterizing the structure evolution are
\begin{enumerate}
\item{}successive formation of structure elements with larger
and larger masses,
\item{} merging of earlier formed, more numerous, low mass
filaments and pancakes
\item{} expansion or compression of matter along filaments
and pancakes
\item{} small scale clustering of compressed matter and 
disruption of structure elements.
\end{enumerate}
The action of these factors was discussed in DD99 (Secs. 4.1, 4.2 
~\&~ 5.2) and it was shown that for the CDM-like initial power 
spectrum and Gaussian initial perturbations it can be approximately 
described by simple analytical expressions. Some of these relations 
were tested and confirmed through the direct comparison with DM 
simulations (DDMT). Here we will briefly summarize these results.

The distribution function of Zel'dovich pancakes with  mass 
(surface density) $m_p$, $N_p(m_p)$, can be written as follows:
$$N_p(m_p)\propto{1\over \sqrt{m_p\langle m_p\rangle}}
\exp\left(-{m_p\over\langle m_p\rangle}\right){\rm erf}\left[\sqrt{
m_p\over\langle m_p\rangle}\right],\eqno(2.4)$$
$$\langle m_p\rangle\propto B^2,$$ 
where the successive merging of low mass pancakes in the course 
of formation of richer pancakes is described by the $erf$-function.
The process of merging decreases the number density of numerous 
low mass structure elements -- both filaments and pancakes -- with 
$m_p\ll\langle m_p\rangle$, but its influence becomes negligible 
for rare, massive, wall-like structure elements, with $m_p\geq\langle 
m_p\rangle$, and rare rich filaments. For the fraction of pancakes 
with $m_{min}\leq m_p\leq m_{max}$ from (2.4) we have:
$$f_p\propto erf^2\left[\sqrt{m_{max}\over
\langle m_p\rangle}\right]-erf^2\left[\sqrt{m_{min}\over
\langle m_p\rangle}\right].\eqno(2.5)$$ 
For rich walls formed presumably at $z\leq$ 1 -- 1.5 with 
$m_{max}\rightarrow\infty$ and $m_{min}\equiv m_w\sim\langle m_p
\rangle$ this relation is simplified and approximately we have 
$$f_w\propto 1-erf^2\left[\sqrt{m_w\over\langle m_p\rangle}\right]
\propto erfc\left[\sqrt{m_w\over\langle m_p\rangle}\right],
						\eqno(2.6)$$ 
what constrains the population of massive pancakes. 
For the subpopulation of poor pancakes similar relation 
$$f_w\propto 1-erf^2\left[\sqrt{m_{min}\over\langle m_p\rangle}
\right]\propto exp\left[-{m_{min}\over\langle m_p\rangle}\right],$$ 
is valid only at higher redshifts $z\geq$ 3.5 -- 4, when 
$m_{max}\gg m_{min}\sim\langle m_p\rangle$, whereas during the 
most interesting period, $z\leq$ 3.5, when  $m_{min}\ll\langle 
m_p\rangle$, the second term in (2.5) can be omitted. 

The discrimination of rich walls and poor pancakes seems 
to be artificial, in some respects, but it allows us to provide 
a closer connection with observed galaxy and simulated DM
distributions at smaller redshifts. Moreover, rich walls 
are certainly strongly nonhomogeneous and disrupted into 
high density clumps and filaments, what essentially decreases 
their covering factor. Due to these peculiarities such walls 
can form a special class of objects. 

The distribution function of filaments with mass (linear density) 
$m_f$ is found to be proportional to 
$$N_f\propto{1\over\langle m_f\rangle}K_0^2\left(\sqrt{m_f\over 4
\langle m_f\rangle}\right), \quad \langle m_f\rangle\propto B^4,
							\eqno(2.7)$$
where $K_0(x)$ is the modified Bessel function (DD99) and 
the fraction of rich filaments, with $m_f\geq \langle m_f\rangle$, 
is 
$$f_f\propto \exp\left(-\sqrt{m_f\over\langle m_f\rangle}\right).
							\eqno(2.8)$$ 
As before, the infall of filaments into rich walls decreases the 
fraction of filaments associated with the network $\propto 
erf[\sqrt{m_p/ \langle m_p\rangle}]\propto B^{-1}$. 

Furthermore, the properties of structure are essentially changed
due to matter expansion and/or compression in transversal directions.
Compression results in transformation of pancakes into filaments and
high density clouds, what decreases their effective surface and,
also, the probability to see such a structure element as an absorber.
Strong  expansion increases the surface of pancakes and the length
of filaments and correspondingly decreases their density. In the
limiting case, low mass pancakes and filaments can be transformed into 
a system of high density clumps or they can even completely disperse. 
The expansion decreases 
also the column density of HI below the observational threshold, 
$N_{HI}^{(thr)}\approx 10^{12}cm^{-2}$, what makes such elements 
invisible as  absorbers. Therefore, the population of structure 
elements identified with absorbers is restricted by the condition 
of their relatively slow expansion or compression in transversal 
directions. Fraction of such low mass pancakes decreases with 
time $\propto erf^2[\sqrt{m_p/\langle m_p\rangle}]$, and 
fraction of such low mass filaments decreases $\propto 
erf[\sqrt{m_p/\langle m_p\rangle}]\propto B^{-1}$. The influence 
of these factors is less important for rare rich walls and 
filaments. 

Due to gravitational instability of compressed matter both pancakes
and filaments are usually disrupted into a system of high density
clouds linked by low density bridges. Such disruption occurs more
rapidly in the central high density parts of structure elements, but, 
probably, its impact is not so important for the extended lower 
density halo. When these processes are taken into account
the final relations become very cumbersome and therefore such 
more detailed treatment is not really justified for when only 
limited set of observed absorbers is available. 

When these factors are taken into account we can only approximately 
describe the expected evolution of walls, filaments, and low mass 
pancakes associated with the absorbers. The comoving linear number 
density of massive walls, $\sigma_w(m)$ with $m\geq m_w$, the comoving 
surface number density of filaments, $\sigma_f(m)$, and the 
comoving number density of pancakes, $\nu_{pan}(m)$, can be 
approximated by simple expressions:
$$\sigma_w\propto \exp\left(-{b_w(m)\over B^2}\right),\eqno(2.9)$$
$$\sigma_f\propto B^{-2}\exp\left(-{b_f(m)\over B^2}\right),
						\eqno(2.10)$$
$$\nu_{pan}\propto erf^2\left({b_1\over B}\right)
\left[erf^2\left({b_2\over B}\right)-erf^2\left({b_3
\over B}\right)\right],				\eqno(2.11)$$ 
where the parameters $b_w, b_f~\&~ b_3(m)$ and the exponential 
terms in (2.9) and (2.10) restrict formation of structure elements 
with masses $>m$, the term $B^{-2}$ in (2.10) describes successive 
infall of filaments into richer walls and the decrease, due to   
expansion or compression, of number of filaments observed 
as absorbers with $N_{HI}\geq N_{HI}^{(thr)}$. The decrease of the  
number density of pancakes is described by the factor $b_1$ in (2.10).

For restricted redshift interval $z\leq$ 3.5 the three parameter 
expression for $\nu_{pan}$ can be approximated by a simpler one 
parameter expression 
$$\nu_{pan}\propto B^{-3}\exp\left(-{b_{pan}(m)\over B^2}\right),
						\eqno(2.12)$$
which  correctly describes the asymptotical behavior of $\nu_{pan}$ 
at small and large redshifts. In the intermediate region it provides
a reasonable precision 
$\sim$ 10 -- 15\%, for $z\leq$ 3.5. The more complicated relation 
(2.11) provides better description of absorbers evolution 
especially at higher redshifts $z\geq$ 3 -- 3.5 and can be used 
for more refined fits. 

High density clouds and galaxies are usually embedded in 
filaments and pancakes and cannot be distinguished as a special 
class of absorbers. These problems were discussed with more details 
in our other  publications DD99, DMRT and DDMT.

\subsection{Model of absorbers evolution}

The relations (2.3), (2.9) - (2.12) cannot give the full description of
observed redshift distribution of absorbers, but they suggest
possible redshift evolution of various types of absorbers and
introduce the function $B(z,\Omega_m)$ as an important characteristic
of such evolution. It is important that the redshift evolution
of free-path between walls, filaments, and low mass pancakes is
expected to be different, and therefore, it might be possible to
single out statistically these three subpopulations of absorbers, 
and to establish correlations between the evolutionary rates and 
other observational characteristics of absorbers.

The DM structure is quite complicated in itself as it is composed 
of several types of structure elements with different evolutionary
histories. Furthermore, properties and evolution of the observed
gaseous and DM structure elements are not identical, because of the
influence of additional factors on the evolution of gaseous component
of DM confined structure elements and, therefore, even more
complicated evolution of absorbers can be expected. The action of 
these factors usually decelerates the evolution of gaseous component 
but, due to very small density of this component, its impact on the 
evolution of DM component is negligible. Thus, the formation of 
weak absorbers within "minivoids" does not accompanied by the formation 
any DM structure elements (Bi \& Davidsen 1997; Zhang et al. 1998; 
Dav\'e et al. 1999). 

The observational restrictions such as the condition $N_{HI}\geq 
N_{HI}^{(thr)}$ constrain also the population of absorbers and 
the observed evolution of absorbers.
Thus, for example, the exponential growth of fraction of massive
high temperature pancakes (in the limiting case, walls similar to the
Great Wall) is not found in the observed distribution of absorbers
(Sec. 4). Moreover, the mean measured $b$ -- parameter of absorbers,
related to the gas temperature, remains constant for the observed
redshifts 2.5 $\leq z \leq$3.3. This means that, possibly, some 
rich DM structure elements are not yet identified in the absorption 
spectra of quasars.  

The redshift dependence of $n_{abs}$ is driven both by the
evolution of spatial characteristics of DM structure elements,
discussed above, and by evolution of their covering factor, which
characterizes the probability of formation of absorption line
at the intersection of line of sight and a structure element.
The covering factor depends on local conditions such as, for 
example, variations of the UV background and activity of nearest 
galaxies. Now it can be also described phenomenologically. 

The important role of UV background is well established after the
discussion in Sargent et al. (1980), but the UV intensity and 
its possible variation with redshift is known only with large 
uncertainty (see, e.g., Haardt \& Madau 1996; Cook, Espey 
\& Carswell 1997). The influence of systematic variations of 
UV radiation is more important at small ($z\leq$ 2) and higher 
($z\geq$ 3 -- 3.5) redshifts where the statistic of absorbers 
is strongly limited. At intermediate redshifts 2.5 $\leq z\leq$ 3.5 
where the main fraction of observed absorbers is situated the 
systematic variations of UV radiation are not so strong, but 
this radiation is probably responsible for the significant 
irregular variations of absorbers density. 

The redshift distribution of observed absorbers can be distorted 
by the formation of artificial caustics in the redshift space (McGill 
1990; Levshakov \& Kegel 1996, 1997). As was shown in DDMT the 
impact of this effect is certainly small at small redshifts. 
At intermediate redshifts and for the usually used CDM-like power 
spectra the formation of artificial caustics is partly suppressed 
due to strong matter concentration within low mass structure 
elements discussed in DD99 and Zhang et al. (1998). This factor 
transforms the continuous matter infall into pancakes to discontinuous 
one, increases the density gradient near pancake boundaries and partly 
prevents the formation of artificial caustics. Moreover, such an 
artificial caustic is actually the preliminary stage of formation 
of a real caustic, and, so, it is rapidly  transformed into a real 
one. This factor can moderately change the discussed redshift 
dependence of absorbers. 

The contribution of artificial caustics and absorbers within "minivoids" 
as well as of short-lived rapidly expanding or contracting pancakes 
is more important at higher redshifts, when weak absorbers dominate. 
These factors generate an essential noise what is usually typical 
for the period preceding the epoch of regular evolution. At such 
redshifts, the available statistics of observed absorbers is very poor, 
and our analysis becomes unreliable. Perhaps, more detailed and 
careful investigations based upon the simulations will allow to
discriminate observationally between this noise and long-lived
structure elements (see, e.g., discussion in Zhang et al. 1998) 
that will essentially improve and simplify the comparison with 
theoretical expectations. 

The more detailed description of structure evolution significantly
increases  the number of fit parameters for limited available 
statistics of observed absorbers, and more detailed description 
accompanied by a further growth of number of fit parameters 
does not seem to be justified. Hence, as the first step, we will 
fit the observed redshift distribution of absorbers to relations 
similar to theoretically expected expressions (2.9)--(2.12), but instead 
of the functions $b(m)$, arbitrary fitting parameters will be 
used. The difference between the observed and expected redshift 
distributions can be attributed to the influence of omitted factors. 
Similar analysis can be repeated with richer data base and/or with 
available numerical simulations.

\subsubsection{Absorption in wall-like elements}

The first term in (2.2), $n_w$, describes absorption associated with 
wall-like elements. Now these elements accumulate $\approx$ 50\%
of galaxies and their possible contribution cannot be rejected
{\it a priori}. The observed mean separation of walls is about
40 -- 60$~h^{-1}$Mpc and the comoving linear number density of
such elements is $\sigma_w(0)\approx$2.5 -- 1.5$\cdot 10^{-2}
h$Mpc$^{-1}$ (LCRS1). These elements are formed during late 
evolutionary stages and can be observed, primarily, at small 
redshifts, $z\leq$1 -- 1.5.

Taking into account expressions (2.3) ~\&~ (2.9) we will approximate
the mean dimensionless linear number density of absorbers associated
with such elements by the two parameter function
$$n_w(z)\approx \kappa_w E(c_w,z),\eqno(2.13)$$
$$\kappa_w = {c\sigma_w(0)\alpha_w\over H_0},\quad
E(c_w,z)=\exp[-c_w(B^{-2}(z)-1)],$$
where the covering factor $\alpha_w<1$, and the parameter $c_w$
characterize probability of formation of an absorption line in a wall
and the period of wall formation, respectively. The observed
wall-like structure elements are strongly disrupted (see, e.g.,
Fig 5 in Ramella et al. 1992) and therefore we can expect that
$\alpha_w\ll$1, $n_w(0)=\kappa_w\approx 60\alpha_w$.

\subsubsection{Absorption in filamentary elements}

The second term in (2.2), $n_f$, describes  absorption in filaments
formed by DM, galaxies and intergalactic gas. The filamentary
component of the structure accumulates $\approx$ 50\% of DM and
galaxies and forms a joint network of structure between wall-like
elements. As was noted above, the mean observed 2D surface density
of galaxy filaments is estimated as $\sigma_f(0)\approx
0.01h^2$Mpc$^{-2}$.

The mean free-path between filaments depends on their surface
density, $\sigma_f$, and their diameters. To find it we consider
the intersection of cylindrical filaments with typical radius of
gaseous halo $R_f$ and a random cylindrical core with radius $r$
and length $L$. The mean number of such intersections can be found
 with standard methods (Kendall \& Moran 1963, Buryak et al. 1994)
as follows:
$$\langle N_{int}\rangle=\pi\sigma_f L(r+R_f).$$
The mean free-path between such cylindrical filaments along a line
of sight can be approximated by
$$\langle l_f\rangle\approx L/\langle N_{int}\rangle|_{r=0} =
(\pi\sigma_f R_f)^{-1}, \eqno(2.14)$$
while the mean length of a line of sight within a filament is
$\approx 2R_f$. 

Taking into account expressions (2.10) we will approximate the
mean dimensionless linear number density of absorbers associated
with filaments by a two parameter function
$$n_f(z)\approx \kappa_f(1+z) B^{-2}(z) E(c_f,z)\eqno(2.15)$$
$$\kappa_f = {c\over H_0}\pi R_{eff}\sigma_f(0),\quad
E(c_f,z) = \exp[-c_f(B^{-2}(z)-1)],$$
$$R_{eff}=\alpha_f\langle R_f\rangle  ={\kappa_f H_0\over
\pi c\sigma_f}\approx 10.6\kappa_f\left({\sigma_f\over
0.01h^2{\rm Mpc}^{-2} }\right)h^{-1}{\rm kpc}.$$
Here $\langle R_f\rangle$ is the mean radius of gaseous halo of a
filament, the covering factor $\alpha_f(z)<1$ characterizes the
probability of formation of an absorption line within a separate
filament, and the factors $1+z$, and $B^{-2}(z)$, describe the expected
variation of surface density of absorbers associated with
filaments due to the general expansion of the universe, merging,
and expansion or compression of filaments. The parameter $c_f$
characterizes the period of formation of the main fraction of
observed filaments.

The mean radius of gaseous halo of filaments, $\langle R_f\rangle$,
can vary with the redshift and along the filament, because the
observed galactic density varies along the filament as well.
This means that our statistical estimate of averaged $\langle
R_f\rangle$ is weighted by the matter distribution along filaments
at different redshifts.

More detailed models of absorbers associated with DM filaments
can be developed. They depend however on many parameters
which cannot be adequately determined from the available
database. Hence, in this paper we will restrict our consideration
to the two parametric model discussed above.

\subsubsection{Absorption in low mass structure elements}

The third term in (2.2), $n_{pan}$, describes absorption
by low mass structure elements. The available information
about properties of this population at small redshifts is very
limited and we can estimate only the mean linear number density
of such pancakes $n_{pan}(0)=\kappa_{pan}$. 
Taking into account expressions (2.12), we will approximate the mean
dimensionless linear density of such absorbers, as
$$n_{pan}(z)\approx \kappa_{pan}(1+z)^2B^{-3}(z)E(c_{pan},z),\eqno(2.16)$$
$$E(c_{pan},z) = \exp[-c_{pan}(B^{-2}(z)-1)].$$
Here the factors $(1+z)^2$, and $B^{-3}(z)$, describe the expected
variation of the density $n_{pan}(z)$, caused by the general
expansion, merging of pancakes, and their compression, and/or
expansion in transversal directions. The parameter $\kappa_{pan}$
is the effective dimensionless free-path between such absorbers at
$z=0$, and $c_{pan}$ characterizes the period of formation of
the main fraction of observed pancakes with
 $N_{HI}\geq N_{HI}^{(thr)}$.

The expressions (2.13), (2.15) ~\&~ (2.16) give probable fitting
relations to the observed evolution of absorbers associated with
various types of structure elements. They allow us to estimate,
in principle, the effective radius of gaseous component of
filaments and the parameters of cosmological model $\Omega_m$
and $\Omega_\Lambda$. All fitting parameters have  clear
interpretation.

Here we do not consider the possible contribution of any high
density clouds, because they are usually embedded in DM filaments
and pancakes. This contribution is small, at least at redshifts
$z\leq$ 3, and cannot be reliably singled out with available
database, but it probably is more important at higher redshifts.

\subsection{Observed linear number density of structure
elements}

The small scale clustering of absorption lines has been reported
in many papers (see, e.g., Cristiani et al. 1995, 1996; Ulmer 1996),
and it is especially important for metal systems. It implies that a
complex of nearby lines can be generated in the same structure
element. As we are interested in  statistics of structure elements
this factor must be taken into account. To do this we will use a
technique developed earlier for the core-sampling method (Buryak
et al. 1994; LCRS1), which allows us to select a Poison-like subsample
of points from a general sample. The redshift dependence of the mean
number density of the subsample characterizes the redshift evolution
of uncorrelated absorbers, which can be identified with separate
structure elements.

This method uses separation of neighboring lines rather than
the lines redshift itself, what attenuates somewhat the influence
of selection effects inherent in individual spectra. If the number
of lines in a sample is $\geq$ 30 -- 50 we can
estimate the parameters of Poisson distribution
with a reasonable precision $\approx$ 10 -- 20\%. Hence, we can use
the available catalogues, what is important for the metal systems,
as number of such systems in individual absorption spectra is
usually small.

The major points of the method can be summarized as follows:
\begin{enumerate}
\item   Separation between lines is characterized by the dimensionless
	comoving distance
	$$\Delta l = H_0\Delta z/H(z).\eqno(2.17)$$
\item   A subsample of lines, in an interval $z-dz< z < z+dz$,
	taken from all spectra under investigation is
 	organized into an `equivalent single field' by combining
	the line separations $\Delta l$ one after the other
	along a line.
\item   Distribution of absorbers obtained in that way
	is assumed to be Poissonian for larger separations.
        1D cluster analysis is used to discriminate the
	Poisson-like subsample among the sample of points, and
        to find the number, and the mean linear number density of
	such points. The theoretical groundwork for such an
        approach was developed by Buryak, Demia\'nski, \&
	Doroshkevich (1991).
\end{enumerate}

Using the standard 1D clustering analysis, the mean number of
Poisson points in the sample, $N_P$, and their mean linear number
density, $n_P$, are found by the maximum likelihood fit to the
relation
$$\ln(N_l) = \ln(N_P) - n_PR_l,\eqno(2.18)$$
where $R_l$ and $N_l$ are the variable linking length and the
related number of clusters, respectively. Note that for a truly
Poisson sample parameters $N_P \& n_P$ are related to the length
of the `equivalent single field', $D_0$, defined by the first and
the farthest point by
$$D_0/N_P = n_P.\eqno(2.19)$$
Thus, the difference between the values $N_P$ and $n_p$
obtained from equation (2.18) and from equation (2.19), and
variations of the actual number of clusters along a straight
line (2.18) give us a measure of error made due to the
difference of the actual from the assumed (Poisson) distribution
for absorbers along the line of sight. To decrease this error we
use an automatic procedure, which finds the optimal range of $R_l$
for the fit to equation (2.18). Both upper and lower limits of $R_l$
were varied to obtain the best fit for the linear number density
of absorbers $n_{abs}(z)$.

As a rule, the precision of this method increases with the
number of lines, $N_{line}(z)$, in the interval $z-dz< z < z+dz$.
For $N_{line}(z) >50$, any random variation does not exceed 10\%, and
the real error is defined by the systematic variations of the sample
under investigation. For smaller $N_{line}(z)$, random variations
become essential. This condition restricts the choice
of optimal interval to $dz\approx$ 0.2 -- 0.25.

\section{ The database}

The present analysis is based on spectra available in the
literature. The list of such objects is given in Table 1.

The distribution of lines over redshift is nonhomogeneous and
majority of lines are concentrated at $z\approx$ 3. Using the
technique described above, we can obtain the linear density of
Poisson subsample of absorbers, at $z\leq 3.2$, with a reasonable
precision of about of 10 -- 20\%. Distribution of absorbers, at
$z\geq 3.2$, is based primarily on the 430 lines of QSO 0000-260
(Lu et al. 1996), and here the statistics of lines is rather poor.
Inclusion of the spectrum of QSO 1033-033 extends the redshift
interval up to $z\approx$ 4.4, but it cannot improve inadequate
representativity of the sample at $z\geq$ 3.2. Samples of poorer
lines, with $\log N_{HI}\leq$ 13, are incomplete at all redshifts.

A few samples of observed metal systems were analyzed. Here we
present results obtained for the richest recent sample (Vanden Berk
et al. 1999). This catalogue contains 901 separations between
lines of neighboring metal systems  from 237 spectra of
quasars at redshifts $0\leq z\leq 3.5$.

\begin{table}
\caption{QSO spectra from the literature}
\label{tbl1}
\begin{tabular}{cccccc}
Name&$z_{em}$&$z_{min}$&$z_{max}$&FWHM&No \\
               &     &   &   &km/s&of lines\\
$1331+170^{1}$ & 2.10&1.7&2.1& 18&~~69\\
$1101-264^{2}$ & 2.15&1.8&2.1&~~9&~~84\\
$1225+317^{3}$ & 2.20&1.7&2.2& 18&159\\
$1946+766^{4}$ & 3.02&2.4&3.0&~~8&461\\
$0636+680^{5}$ & 3.17&2.5&3.0&~~8&313\\
$0302-003^{5}$ & 3.29&2.6&3.1&~~8&266\\
$0956+122^{5}$ & 3.30&2.6&3.1&~~8&256\\
$0014+813^{5}$ & 3.41&2.7&3.2&~~8&262\\
$0000-260^{6}$ & 4.11&3.4&4.1&~~7&431\\
$2126-158^{7}$ & 3.26&2.9&3.2& 11&130\\
$0055-259^{8}$ & 3.66&2.9&3.1& 14&313\\
$1700+642^{9}$ & 2.72&2.1&2.7& 15&~~85\\
$2206-199^{10}$& 2.56&2.1&2.6& 11&101\\
$1033-033^{11}$& 4.50&3.7&4.4& 18&299\\
\vspace{0.15cm}
\end{tabular}

1. Kulkarni et al. (1996),
2. Carswell et al. (1991),
3. Khare et al. (1997),
4. Kirkman \& Tytler (1997),
5. Hu  et al., (1995),
6. Lu et al. (1996),
7. Giallongo et al. (1993),
8. Cristiani et al. (1995),
9. Rodriguez et al. (1995),
10. Rauch et al. (1993),
11. Williger et al. (1994),
\end{table}

The published Ly-$\alpha$ lines with different $N_{HI}$ were
organized into six samples listed in Table 2. All samples were
supplemented by the HST data (Bahcall et al. 1993, 1996; Jannuzi
et al. 1998). The samples with $N_{HI}\geq
13.8$, are associated mainly with the filamentary component of
the structure, whereas samples with $N_{HI}\geq 13$, and
$N_{HI}\geq 12$, are associated predominantly with the poor
pancakes.

\begin{table}
\caption{Samples of absorbers.}
\label{tbl5}
\begin{tabular}{cccccc}
sample    &$N_{QSO}$&$log N_{HI}$&$N_{lines}$&$N_{lines}^{HST}$&$W_{HST}$\\
$Q_{14}^{14} $& 14      &14~~       &~~686      &~~590  &0.5\\
$Q_{14}^{138}$& 14      &13.8       &~~971      &~~590  &0.5\\
$Q_{14}^{13} $& 14      &13.0       & 2351      &~~933  &0.25\\
$Q_{12}^{13} $& 12      &13.0       & 1986      &~~933  &0.25\\
$Q_{14}^{12} $& 14      &12.0       & 3177      & 1000  &0.\\
$Q_{12}^{12} $& 12      &12.0       & 2780      & 1000  & 0.\\
\end{tabular}
\end{table}

\section{Statistical analysis of structure evolution}

In this section the main results are presented for the
redshift distribution of both Ly-$\alpha$ lines and metal
systems. They confirm that the redshift distribution of
Ly-$\alpha$ lines is a superposition of several populations,
with different evolutionary histories (Bahcall et al. 1996).
Using the method and the fitting relations discussed in Sec.
2 for the analysis of samples with different $N_{HI}$, we can
roughly discriminate two populations of absorbers, and describe
their evolution. Our estimates use essentially the HST
data for $z\leq$ 1.5.

For three cosmological models the distribution of absorbers was
fitted to the expression
$$n_{abs} = {\kappa_f(1+z)\over B^2(z)}E(c_f,z)+
{\kappa_{pan}(1+z)^2\over B^3(z)}E(c_{pan},z)\eqno(4.1)$$
$$E(c,z)=\exp[-c\cdot (B^{-2}(z)-1)].$$
The function $B(z)$ was introduced in (2.3).
The main results are plotted in Figs. 1 -- 5, and the best fit
parameters, $\kappa_f, ~c_f, ~\kappa_{pan}$, and
$$R_{eff} = {\kappa_f H_0\over
\pi c\sigma_f}\approx 10.6\kappa_f\left({\sigma_f\over
0.01h^2{\rm Mpc}^{-2} }\right)h^{-1}{\rm kpc},\eqno(4.2)$$
are listed in Tables 3 -- 5. For samples of richer absorbers
$c_{pan}$=0, whereas for samples of poorer absorbers
$c_{pan}\approx 0.05\pm 0.05$ was found with large errors and,
so, it is omitted in Table 4.

\subsection{The redshift distribution of metal system and
stronger HI lines}

It can be expected that the redshift distribution of metal
systems and stronger Ly-$\alpha$ lines is connected with
the evolution of richer structure elements, which are seen
in galaxy surveys as filaments and walls. To examine this
hypothesis the function $n_{abs}(z)$ was found for the
sample of metal systems, and the samples of strongest
Ly-$\alpha$ lines $Q_{14}^{14}$, and $Q_{14}^{138}$. The analysis
shows that in all cases no more than 10 -- 15\% of all lines can
be attributed to the massive wall-like elements, what is consistent
with the large separation of walls ($\geq 50h^{-1}$Mpc) and their 
strong disruption into a system of high density clouds, what 
decreases their covering factor. 

For all considered cosmological models the redshift distributions
of Ly-$\alpha$ absorbers, for samples $Q_{14}^{14}$, and
$Q_{14}^{138}$, are well fitted, at $z\leq$ 3, to the one parameter
function (4.1) with $c_f=0$, and $\kappa_{pan}=0$.
The redshift distribution of metal systems is well fitted
to the two parameter function (4.1) with $\kappa_{pan}=0$. The
best parameters of these fits are listed in Table 3. They show
that metal systems are  predominantly formed within inner regions of
gaseous halos of filaments, whose size, $R_{eff}^{met}$, is
about half the size of hydrogen halos, $R_{eff}^{HI}$. Both
sizes are consistent with those  observed directly at small
$z$: $R_{HI}\sim$ 100 -- 150$h^{-1}$kpc (Lanzetta et al. 1995),
and $R_{met}\sim$ 40 -- 50$h^{-1}$kpc (see, e.g., Le Brune et
al. 1996), and to the expected size of DM halo (Bahcall et
al. 1996). The precision of estimates of $R_{eff}$ depends on
the representativity of the sample used.

\begin{figure}
\centering
\epsfxsize=8 cm
\epsfbox{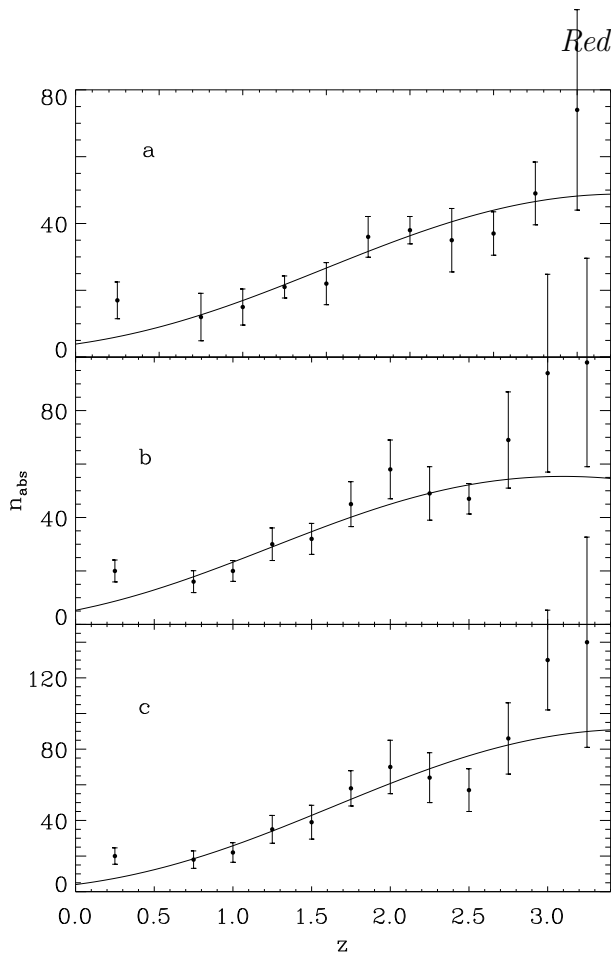}
\vspace{0.75cm}
\caption{$n_{abs}$ vs. $z$ for metal systems  for cosmological 
models with $\Omega_m =0.3,~~ \Omega_\Lambda = 0.7$ (a), 
$\Omega_m =0.5,~~\Omega_\Lambda = 0$ (b), and $\Omega_m =1$ (c).
}
\end{figure}

\begin{figure}
\centering
\epsfxsize=8 cm
\epsfbox{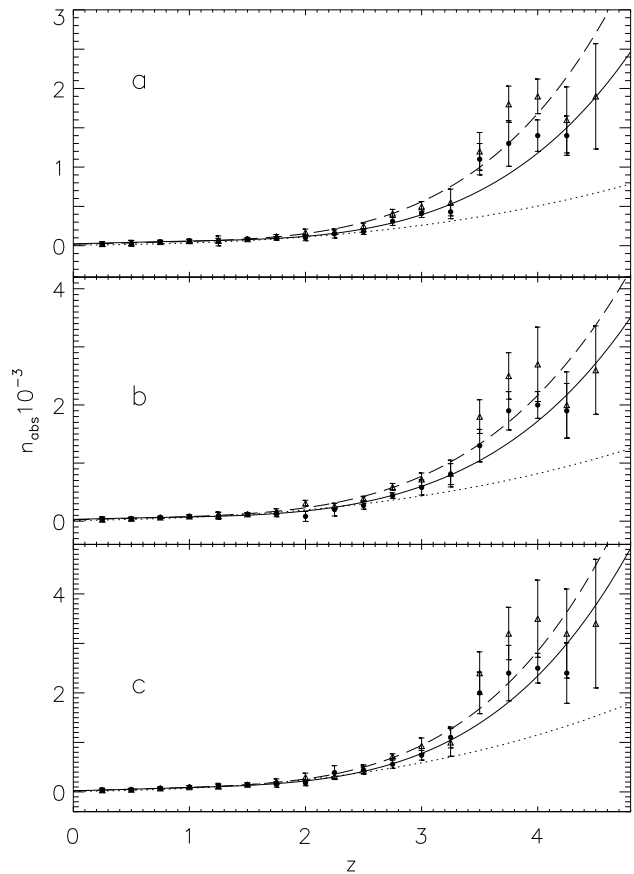}
\vspace{0.75cm}
\caption{The redshift distribution of absorbers for samples
$Q_{14}^{14}$ (points \& solid lines) and $Q_{14}^{138}$
(triangles \& long dashed lines), for cosmological models with
$\Omega_m =0.3,~~ \Omega_\Lambda = 0.7$ (a), $\Omega_m =0.5,~~
\Omega_\Lambda = 0$ (b), and $\Omega_m =1$ (d). Dot lines
show the best two parameters fit for the filamentary component
only at $z\leq$3.
}
\end{figure}

\begin{table}
\caption{Fit parameters for the filamentary component at $z\leq$3.}
\label{tbl3}
\begin{tabular}{ccccc}
$\Omega_m$&$\Omega_\Lambda$&$R_{eff}^{met}$&$c_f^{met}$&
$R_{eff}^{HI}$\\
       &    &$h^{-1}kpc$ &           &$h^{-1}kpc$\\
  1.0  & 0~~& $40\pm 1.5$&$0.07\pm0.03$&$~~98\pm2.4$\\
  0.5  & 0~~& $52\pm 1.6$&$0.15\pm0.04$&$ 118\pm3.3$\\
  0.3  & 0.7& $38\pm 1.7$&$0.13\pm0.04$&$~~69\pm3.5$\\
\end{tabular}
\end{table}

\begin{figure}
\centering
\epsfxsize=8 cm
\epsfbox{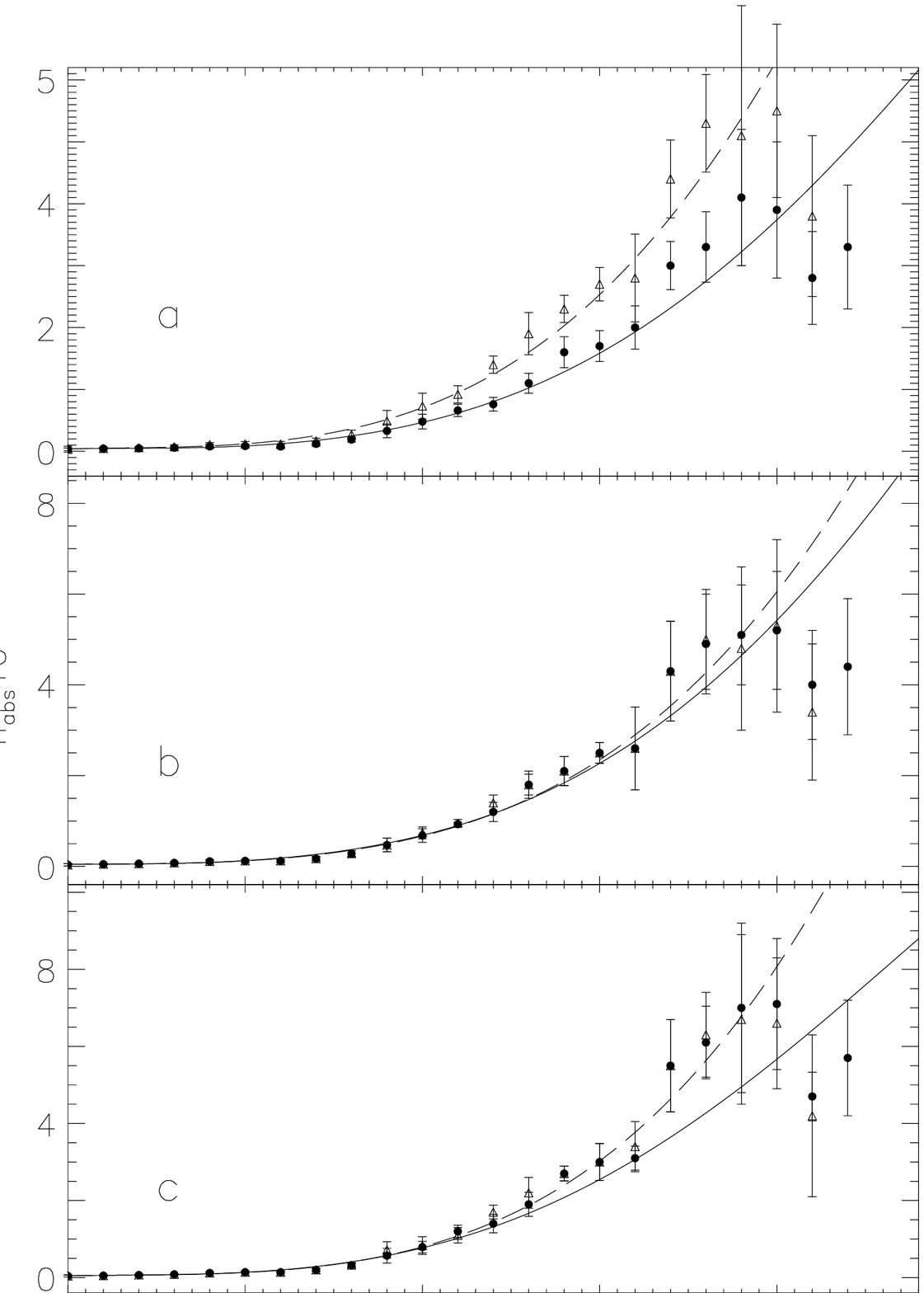}
\vspace{0.75cm}
\caption{The redshift distribution of absorbers for
the sample $Q_{14}^{13}$ (points \& solid lines) and
the sample $Q_{12}^{13}$ (triangles \& long dashed lines),
for cosmological models with $\Omega_m =0.3,~~ \Omega_\Lambda = 0.7$ (a),
$\Omega_m =0.5,~~\Omega_\Lambda = 0$ (b), and
$\Omega_m =1$ (c).
}
\end{figure}

\begin{figure}
\centering
\epsfxsize=8 cm
\epsfbox{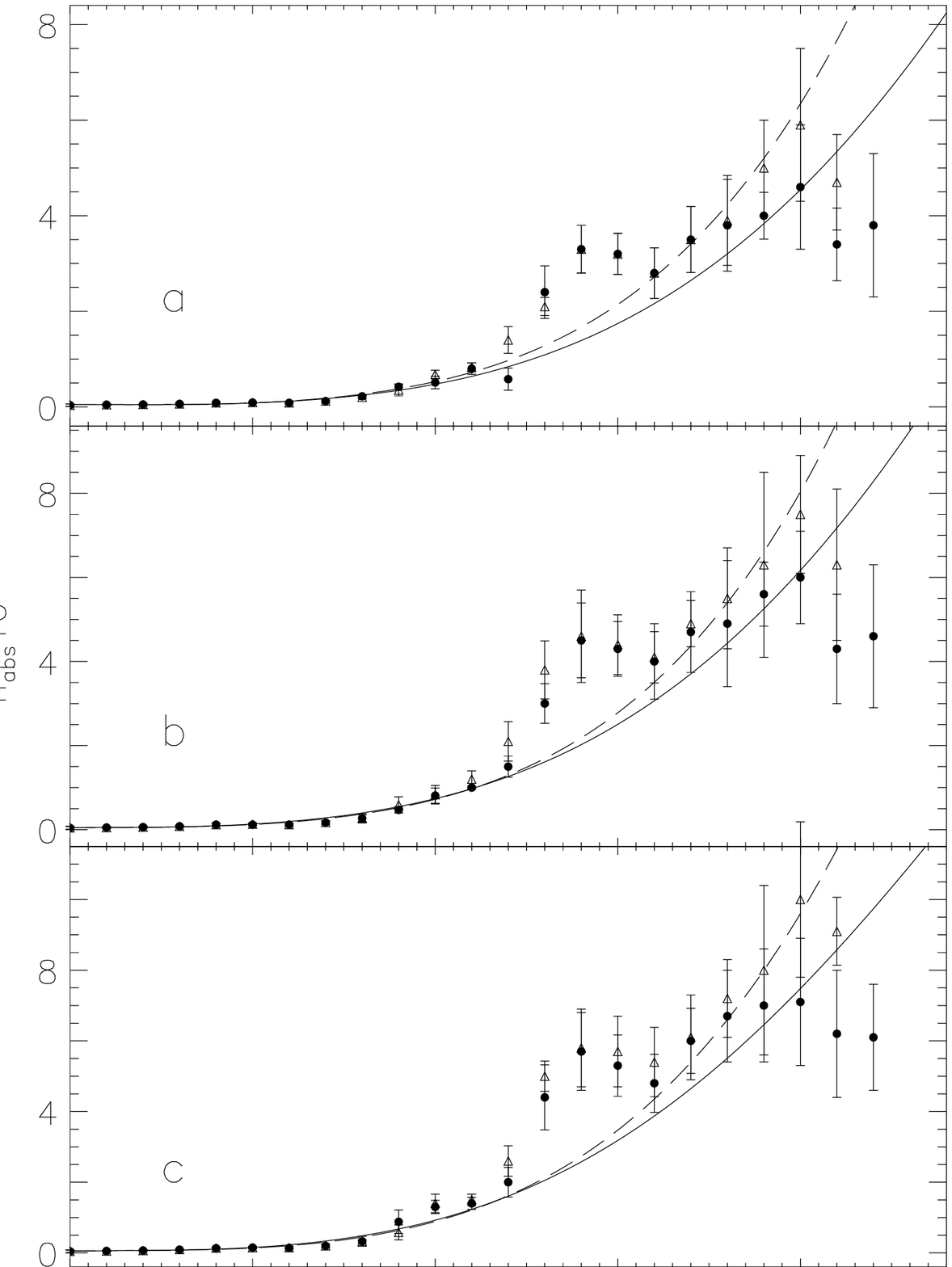}
\vspace{0.75cm}
\caption{The redshift distribution of absorbers for
the sample $Q_{14}^{12}$ (points \& solid lines) and
the sample $Q_{12}^{12}$ (triangles \& long dashed lines),
for cosmological models with $\Omega_m =0.3,~~ \Omega_\Lambda = 0.7$ (a),
$\Omega_m =0.5,~~\Omega_\Lambda = 0$ (b), and
$\Omega_m =1$ (c).
}
\end{figure}

These results suggest that the redshift distribution, and
the column density of absorbers are strongly correlated, and
corroborate  the identification of strong HI absorbers, and metal
systems, with the filamentary component of the structure of
the universe. They show that such filaments were probably
formed at redshifts $z\geq$ 2, but their enrichment by
metals could occur later, at redshifts $z\approx$ 2 and
smaller.

At redshifts $z\geq$ 3, the situation becomes more complicated,
and the observed absorber distribution is no longer described
by the one parameter fit plotted in Fig. 2 by dotted lines.
To obtain a reasonable description, even for the redshift
distribution of stronger absorbers, at $z\geq$ 3, the three
parameter fit (4.1) with $c_{pan}=0$ must be used. This 
indicates that at $z\geq$ 3 progressively increasing part 
of stronger absorbers must be assigned to pancakes. 

The best parameters of these fits are listed in Table 4, and
fitting functions $n_{abs}(z)$ are plotted in Fig. 2 by solid
and long dashed lines. These estimates of $c_f$ and $R_{eff}$
crucially depend on the region of redshifts 2.5 $\leq z\leq$ 3.3, 
and could be
sensitive to the possible evolution of filaments. They
show that the formation of such filaments described by the
exponential term in (4.1) could occur at redshifts $z\geq$ 1,
while later the surface density of filaments 
progressively decreases. The effective radii listed in
Table 4 exceed by about  2 -- 3 times those listed in Table
3 for smaller redshifts. This fact can be attributed to the
progressive compression, small scale disruption, and relaxation
of dark matter within and around filaments, which is accompanied 
by the progressive dissipation of gaseous halo of filaments. 
The observed evolution can also be partly assigned to the 
variations of background UV radiation with redshift. 

At redshifts $z\geq$ 2 -- 2.5, more and more essential fraction of
strong absorbers is associated with rich pancakes, and, at $z\geq$ 3.5,
the pancakes become dominant, even for this population. This fact
probably indicates the period of more rapid transformation of rich
pancakes into filaments. 

This conclusion is based on poor statistics of absorbers, at
$z\geq$ 3, and should be tested with more representative sample
of absorbers and with simulations.

For our samples the parameters of standard fit to
$$dN/dz = N_0(1+z)^{\gamma_z},\eqno(4.3)$$
are:
$$N_0^{met} = 6.8\pm 0.5,\quad \gamma_z^{met} = 0.35\pm 0.25,$$
$$N_0^{HI} = 12.7\pm 2,\quad \gamma_z^{HI} = 1.5\pm 0.3,\eqno(4.4)$$
with a weak dependence on the considered cosmological models.

\subsection{Redshift distribution of weaker Ly-$\alpha$ lines}

The samples of weaker absorbers are rich enough but, even so,
the main fraction of observed absorbers is concentrated near
$z\sim$ 2.5 -- 3.3. In Figs. 3 \& 4 the functions $n_{abs}(z)$ are
plotted, for the samples $Q_{14}^{13}$, $Q_{12}^{13}$, and $Q_{14}^{12}$,
$Q_{12}^{12}$, together with the four parameters fit (4.1).
Both samples of poor absorbers with $\log N_{HI}\leq 13$ 
are incomplete. The best fit parameters $\kappa_f$, $c_{f}$, 
$\kappa_{pan}$, and $c_{pan}$ listed in Table 4 are weakly 
sensitive to the used cosmological models.

These results show that, at redshifts $z\leq$ 1.5 -- 2, an
essential fraction of weaker absorbers can be associated with the
periphery of filaments and/or with possible membranes and bridges
between branches of filaments, what is consistent with the
expectations of Fernandes-Soto et al. (1996). But at high
redshifts, the main fraction of poorer absorbers is certainly
identified with the population of low mass pancakes and, for 
$N_{HI}\leq 10^{13}cm^{-2}$, with absorbers formed within expanded 
regions. The small
value of $c_{pan}\sim$ 0.03 -- 0.05 suggests that these absorbers
were formed at $z\approx$ 4 -- 5. These estimates are sensitive
to the distribution of absorbers at $z\geq$ 3, where the observed
samples are not sufficiently representative. For these populations
the random overlapping of poor absorbers in the redshift space,
discussed by McGill (1990) and Levshakov \& Kegel (1998), formation 
of absorbers within "minivoids" (Zhang et al. 1998; Dav\'e et al. 
1999) as well as the variations of UV background can distort the 
observed mass and temperature distribution of absorbers. As was 
noted in Sec. 2.1.2, at these redshifts, the more complicated three 
parameters function (2.11) provides better fit of observed distribution 
of absorbers. 

The essential variations of measured linear density, $n_{abs}$,
with respect to the smooth fitting functions are seen in Figs. 3
\& 4 at $z\approx$ 2.5 -- 3. They can be caused, in part, by
the poor statistics of absorbers, at $z\approx$ 2.5, and $z\geq$ 3.5.
If the variations seen at $z\approx$ 3 are real they can be
connected with similar results obtained for the population of
stronger absorbers and, thus, can indicate the period of fast
transformation of pancakes into filaments. They can also be caused
by influence of local factors and, first of all, by variations of 
background UV radiation.

The parameters of standard fit (4.3) are:
$$N_0^{pan} = 13.8\pm 0.6,\quad \gamma_z^{pan} = 2.4\pm 0.2,~~
{\rm for}~~Q_{12}^{13},\eqno(4.5)$$
$$N_0^{pan} = 11.8\pm 0.6,\quad \gamma_z^{pan} = 2.75\pm 0.2,~~
{\rm for}~~Q_{12}^{12},\eqno(4.6)$$
with a weak dependence on the parameters of considered cosmological 
models. The power index $\gamma_z^{pan}$ is close to the value found 
earlier (see, e.g., Carswell 1995; Cristiani et al. 1996).

\subsection{Redshift distribution of Ly-$\alpha$ lines
with $b\geq$ 40km/s and $b\leq$ 20km/s }

The same technique can be applied to single out other subpopulations
of absorbers such as subpopulations with lower Doppler parameter $b$, $b\leq$ 20km/s, and
larger, $b\geq$ 40km/s. The main results of this
analysis are plotted in Fig. 5, and the best fitting parameters are
listed in Table 5. Unfortunately, they cannot be complemented by
HST data at smaller redshifts.

\begin{table}
\caption{Fitting parameters of redshifts  distribution of absorbers
for three cosmological models.}
\label{tbl4}
\begin{tabular}{ccc cc} 
sample&$\kappa_f$&$R_{eff}$&$c_f$&$\kappa_{pan}$\cr
         &              &$100h^{-1}$kpc&     &                 \cr
\hline
\multicolumn{4}{c}{$\Omega_m=1,~~ \Omega_\Lambda=0$}\cr
$Q_{14}^{14}$ &$22\pm1.3$&$2.2\pm0.1$&$0.34\pm 0.14$&$~~0.75\pm 0.12$\cr
$Q_{14}^{138}$&$21\pm1.2$&$2.1\pm0.1$&$0.33\pm 0.15$&$~~0.91\pm 0.11$\cr
$Q_{14}^{13} $&$43\pm1.5$&$4.3\pm0.1$&$1.0~~\pm 0.20$&$~~4.2\pm 0.11$\cr
$Q_{12}^{13} $&$43\pm1.3$&$4.3\pm0.1$&$1.0~~\pm 0.20$&$~~3.7\pm 0.11$\cr
$Q_{14}^{12} $&$45\pm1.6$&$4.4\pm0.1$&$1.1~~\pm 0.22$&$~~4.8\pm 0.11$\cr
$Q_{12}^{12} $&$49\pm2.1$&$4.8\pm0.2$&$1.1~~\pm 0.21$&$~~4.0\pm 0.09$\cr
\hline
\multicolumn{4}{c}{$\Omega_m=0.5,~~ \Omega_\Lambda=0$}\cr
$Q_{14}^{14} $&$30\pm1.5$&$2.9\pm0.2$&$0.66\pm 0.21$&$~~1.2\pm 0.11$\cr
$Q_{14}^{138}$&$23\pm1.3$&$2.3\pm0.1$&$0.54\pm 0.22$&$~~1.6\pm 0.14$\cr
$Q_{14}^{13} $&$39\pm1.5$&$3.8\pm0.2$&$1.82\pm 0.37$&$~~6.4\pm 0.12$\cr
$Q_{12}^{13} $&$39\pm1.4$&$3.8\pm0.2$&$1.62\pm 0.35$&$~~5.8\pm 0.15$\cr
$Q_{14}^{12} $&$44\pm1.6$&$4.4\pm0.6$&$1.93\pm 0.33$&$~~6.7\pm 0.10$\cr
$Q_{12}^{12} $&$38\pm2.0$&$3.8\pm0.2$&$1.63\pm 0.42$&$~~5.6\pm 0.15$\cr
\hline
\multicolumn{4}{c}{$\Omega_m=0.3,~~ \Omega_\Lambda=0.7$}\cr
$Q_{14}^{14} $&$23\pm1.2$&$2.2\pm0.1$&$0.6\pm 0.16$&$~~0.8\pm 0.12$\cr
$Q_{14}^{138}$&$18\pm1.0$&$1.7\pm0.1$&$0.6\pm 0.18$&$~~1.1\pm 0.11$\cr
$Q_{14}^{13} $&$37\pm1.8$&$3.6\pm0.2$&$1.8\pm 0.40$&$~~4.7\pm 0.12$\cr
$Q_{12}^{13} $&$37\pm1.3$&$3.6\pm0.1$&$1.1\pm 0.26$&$~~4.1\pm 0.12$\cr
$Q_{14}^{12} $&$41\pm1.4$&$4.0\pm0.1$&$2.0\pm 0.33$&$~~4.4\pm 0.09$\cr
$Q_{12}^{12} $&$42\pm2.1$&$4.2\pm0.2$&$2.2\pm 0.35$&$~~4.4\pm 0.11$\cr
\hline
\end{tabular}
\end{table}

\begin{table}
\caption{Fit parameters for the subsamples of absorbers
$Q_{14}^{12}$ with $b\geq$ 40km/s and $b\leq$ 20km/s}
\label{tbl5}
\begin{tabular}{ccccc}
$\Omega_m$&$\Omega_\Lambda$&$\kappa_f$&
$\kappa_{pan}$&$10^2\cdot c_{pan}$\\
\hline
\multicolumn{4}{c}{$b>40$km/s}\cr
  1.0  & 0~~&0.0&$~~2.4\pm 2.4$&$5.4\pm 2.2$ \\
  0.5  & 0~~&0.0&$~~1.9\pm 2.0$&$3.6\pm 6.0$\\
  0.3  & 0.7&0.0&$~~2.4\pm 3.1$&$7.7\pm 4.1$\\
\hline
\multicolumn{4}{c}{$b<20$km/s}\cr
  1.0  & 0~~&$11\pm~~4.2$&$0.12\pm0.1$&$1.0\pm 0.5$ \\
  0.5  & 0~~&$16\pm10~~ $&0.0 & 0.0  \\
  0.3  & 0.7&$10\pm~~4.6$&0.0 &0.0\\
\hline
\end{tabular}
\end{table}

These subpopulations are not so representative and at $z\leq$ 4
they contain only 640 lines with $b\leq$ 20km/s and 727 lines with
$b\geq$ 40km/s. This is the main source of the large errors of
fitting parameters listed in Table 5. Nonetheless, these results
demonstrate that, if the subpopulation of hot absorbers with
$b\geq$ 40km/s can be associated with pancakes, then the redshift
distribution of cold absorbers with $b\leq$ 20km/s is similar to
that found for filamentary component of the structure.

\begin{figure}
\centering
\epsfxsize=8 cm
\epsfbox{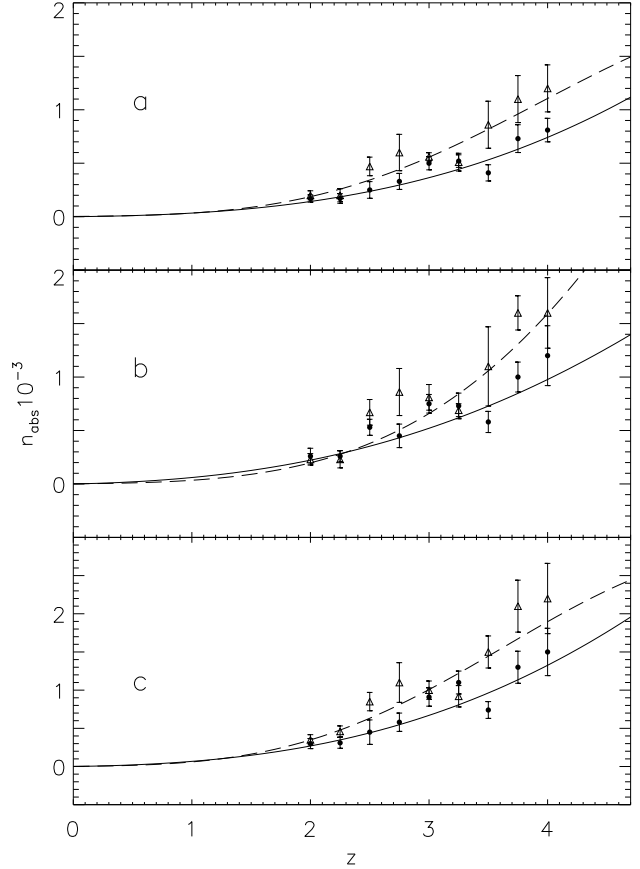}
\vspace{0.75cm}
\caption{The redshift distribution of 640 lines with $b\leq 20$km/s,
(points, solid lines), and 727 lines with $b\geq 40$km/s
for the sample $Q_{14}^{12}$, for three cosmological models
with $\Omega_m=0.3, \Omega_\Lambda=0.7$, (a),
,$\Omega_m=0.5, \Omega_\Lambda=0$, (b), and $\Omega_m=1$, (c).
}
\end{figure}

\section{Summary and Discussion.}

In this paper the observed redshift distributions of Ly-$\alpha$
lines and metal systems are analyzed and interpreted in the
framework of theoretical model of DM structure evolution (DD99).
The observed evolution of absorbers depends on several random
factors and does not trace directly the evolution of the DM component.
Our results show however that the observed redshift distributions
of absorbers can be reasonably described even by the discussed model.

The main results of our analysis can be summarized as follows:
\begin{enumerate}
\item  The different redshift distributions of stronger and
	weaker absorbers favor the existence of two
	distinct subpopulations of Ly-$\alpha$ absorbers:
	a rapidly evolving subpopulation of weaker
	absorbers, that dominates at high redshifts, and more slowly
	evolving subpopulation of stronger absorbers, that dominates
	at low redshifts (Bahcall et al. 1996). This result coincides
	with theoretical expectations.
\item  The expected evolution of filamentary component of
	structure describes adequately the redshift distribution of
	metal systems, and stronger Ly-$\alpha$ lines, at
	$z\leq$ 3. This strong correlation of evolutionary rate
	and the column density of absorbers suggests possible
	identification of stronger absorbers with the filamentary
	component of observed galaxy distribution.
\item The subpopulation of colder absorbers, with $b\leq$ 20km/s, can
	be probably related with the filamentary component of
	the structure.
\item  The rapid evolution of the linear number density of weaker
	absorbers, $n_{abs}$, can be naturally explained by the model
	of separate DM confined pancakes that are  merging and expand or                contract in transversal direction, as it is
 	described by the theoretical model discussed above. At
	$z\leq$ 2 -- 2.5,  disruption of the compressed matter can
	essentially accelerate the evolution of this subpopulation.
\item The observed redshift distribution of absorbers depends on
 	the influence of irregular local factors, such as
	variations of background UV radiation, and shock heating of
	gas produced by activity of galaxies. This impact is seen
	as irregular variations of the observed linear number
	density of absorbers around the smooth fitting curves.
\item  The obtained estimates of $c_f$ can be interpreted as a
	probable pick of the rate of formation of filaments at $z\sim$ 2
	-- 2.5 whereas their enrichment by the metals can occur
	at $z\sim$ 2 and smaller. The small values of $c_{pan}$,
	found for all subpopulations under investigation, show that
	the main fraction of observed poor absorbers can be
	associated with DM structure elements formed at
	redshifts $z\geq$ 5.
\end{enumerate}

The main quantitative characteristics of redshift distribution of
absorbers, discussed above, admit natural interpretation and coincide
with published estimates (see, e.g., Hu et al. 1995; Lanzetta et al.
1995; Cristiani 1995, 1996; Le Brune 1996; Bahcall 1996). They are
roughly consistent with the expected evolution of DM structure and
can be, in principle, used to test and to discriminate different
cosmological models. Now the best fit is obtained for the model
with $\Omega_m$= 0.3, $\Omega_\Lambda$=0.7, but the more representative
database at redshifts $z\sim$ 2 -- 2.5 and $z\geq$ 3 is required
for more reliable discrimination of cosmological models.

On the other hand, the discussed rough model of structure 
evolution ignores many details, such as, for example, the continual 
variation of morphological characteristics of structure elements, 
and does not describe the  disruption of structure. This model could be
improved, but then it is inevitable to increase 
the number of fit parameters and, so, it will be justified only  
when richer sets of observed absorbers, especially 
at smaller and higher redshifts become available. This comment applies also  
to the more detailed description of expected variations 
of UV radiation and possible contribution of artificial caustics. 
Application of this approach to available simulations could test 
and improve it as well.

The wall-like structure elements and large under dense
regions found in distribution of galaxies at small redshifts
cannot yet be reliably identified with the available database.
The large scale modulation of redshift distribution of
Ly-$\alpha$ lines found by Cristiani et al. (1997) and
strong nonhomogeneities, found recently at $z\leq$ 2, by
Williger et al. (1996), Quashnock et al. (1996, 1998), and
Connolly et al. (1997) could be probably attributed to such
rare extremely rich structure elements. Appearance of such 
elements is not forbidden at any redshifts, but it can be expected 
that their number will rapidly decreases for larger redshifts. 

\subsection*{Acknowledgments}
We are grateful to Dan Vanden Berk for providing the 
new list of observed metal systems. 
This paper was supported in part by Denmark's
Grundforskningsfond through its support for an establishment of
Theoretical Astrophysics Center, by the Polish State Committee for Scientific
Research grant Nr. 2-P03D-014-17,  and the grant INTAS-93-68.  AGD and VIT
also wish to acknowledge support
from the Center of Cosmo-Particle Physics in  Moscow.
Furthermore, we wish to thank the anonymous
referee for many useful comments.

\end{document}